\documentclass[a4paper,12pt]{article}
\usepackage{amsmath}
\usepackage{graphicx}
\usepackage{amssymb}
\usepackage{subcaption}
\usepackage{cite}
\usepackage[hidelinks]{hyperref}
\usepackage[font=footnotesize,labelfont=bf]{caption}
\usepackage{lineno}

\hypersetup{
  colorlinks = true,     
  urlcolor = blue,       
  linkcolor = blue,      
  citecolor = red         
}

\newcommand{\R}{\mathcal{R}}

\begin{document}

\title{Quantum Information and the Mind-Body Problem}

\author{M.P. Vaughan\\ University of Essex}

\date{}

\maketitle

\abstract{\footnotesize The mind-body problem is reviewed in the context of a non-technical account of quantum theory. The importance of clearly defining: `what is physical?' is highlighted, since only then can we give meaning to the concept `non-physical'. Physicality is defined in terms of interaction, which is in turn defined to be a correlated exchange of information. This is asserted to be the basis of any meaningful concept of epistemology. Hence, it is argued that a non-physical entity can not `know' anything about the world. Information transfer is then discussed in terms of quantum entanglement and an argument for our perception of time is presented. It is then contended that the notion of `mind' may be meaningfully discussed in the context of a quantum theoretic framework.}

\tableofcontents

\section{Introduction}

Modern quantum theory has radically redefined the way those familiar with it see the world. Perhaps one of the greatest challenges to the student of quantum physics lies in the reconciliation of one's perceptions of the physical world with the counter-intuitive concepts of the theory. We perceive the physical world as consisting of objects, existing in space, independently of our own subjective reality with well-defined locations and structures. Understanding how this `classical' reality emerges from quantum theory has been an ongoing challenge for physicists and quantitative progress has only been made relatively recently, particularly in the growing field of decoherence theory~\cite{kubler1973dynamics, zurek1981pointer, zurek1982environment, schlosshauer2007decoherence}.

On the other hand, the world of our senses is very well described by \emph{classical physics}, which for present purposes we may take to mean Euclidean geometry and Newtonian mechanics. Euclid's axiomatic description of space gives us an extremely accurate model of the spatial relations between these objects, whilst Newtonian mechanics refines our intuitive concepts of the dynamics of corporeal entities into an elegant and well-established mathematical theory.

So persuasive are our perceptions and their mathematical refinement in the form of classical physics that, for many of us, these constitute our fundamental concepts of what physical reality is. That is, we conclude that the physical world consists of objects existing in an external space. Comparing this to our own internal, subjective experiences, it then appears clear that our mental realities appear to be something radically different. We are not (many people believe) governed completely by deterministic physical laws, nor does the holistic experience of our phenomenal reality fit into the cold confines of inert, physical objects. And so, the mind-body problem is born, in which we divide the world into two types of entities: mental and physical.

As a contemporary of Galileo and a major contributor to modern science himself, Descartes was no doubt aware of the emerging formalism of classical mechanics, along with its deterministic description of Nature. It may well have been a sense of unease that such laws purported to describe and govern human experience that spurred him to develop his dualistic description of mind and body. However, his notions of what constituted the physical component of the world would have been guided, at best, by the insight of a classical physical description. The radically different concepts of quantum mechanics would have been entirely alien to him and, at such a time, would have been in direct contradiction to his common-sense notions of the physical world. Such ideas would simply not have figured in his description of the `physical'.

The purpose of this paper is to define and explore what we actually mean by the word `physical' and how that relates to modern quantum theory. Our contention is that this is fundamentally related to the communication of information, which is relevant to both `minds' and `bodies'. 

We begin our discussion by defining what we mean by our terms in Section~\ref{sec:physical} and discussing these concepts in fairly broad terms, before exploring quantum theory in more detail in Sections~\ref{sec:quantum} and \ref{sec:quantum_information}. This latter section specifically focuses on what quantum theory has to say about the process of information transfer. Here, we should clarify that by `quantum information theory', we mean the quantum-mechanical generalisation of Shannon's information theory~\cite{shannon1948mathematical,shannon2001mathematical}. 

Armed with the insights of these sections, we then address the nature of mental phenomena in Section~\ref{sec:quantum_minds}, arguing that there is nothing in our qualitative descriptions of these that necessitates metaphysical explanation. Quantum theory is rich enough to meaningfully discuss such phenomena. We conclude this section with a discussion of the conservation of information and its consequences.

\section{What is `physical'?}\label{sec:physical}
\subsection{Definitions}\label{sec:definitions}

The problems with Cartesian dualism are well-known and centre on the problem of \emph{interaction}. How could a `mind', which exists independently of physical laws, affect a `body', which is entirely determined by those laws, or vice versa? The problem is even more clearly delineated when we stop to think what we actually mean by the concept `physical'. As suggested above, this notion is heavily prejudiced by our everyday perceptions. However, as already alluded to, these everyday notions find little traction in modern quantum physics - as we shall presently discuss. For the present, let us cut to the chase and offer a definition of what the word `physical' means that is consistent with our contemporary scientific understanding:

\begin{itemize}
\item [(1)] An entity is \emph{physical} if and only if it can \emph{interact} with another entity.
\item [(2)] An \emph{interaction} is defined as a correlated exchange of \emph{information}.
\item [(3)] Any physical entity may be quantitatively described in terms of a \emph{state}.
\item [(4)] The \emph{state} constitutes the \emph{informational content} of an entity.
\end{itemize}

Whilst we are about it, we shall also establish the basis for what we might call \emph{quantum epistemology} by defining the word `\emph{know}' to mean `\emph{has information about}'. Here, we are not speaking of complex systems of belief but quite simply how one entity can `know' anything about another. Specifically, we are asserting that information transfer is a \emph{necessary} condition for `knowing' something but not necessarily a \emph{sufficient} one.

In fact, the usage we are implying here is synonymous with the concept of interaction, so we shall explicitly assert:

\begin{itemize}
\item [(5)] An entity can have knowledge of another entity if and only if there is a transfer of information from one to the other.
\end{itemize}

\noindent The concepts of `information' and `correlation' will be discussed in more detail later in the context of quantum information theory. In order to avoid misinterpretation, it is important to interpret the word `information' in the way it is defined in later sections. Specifically, in the context of quantum mechanical mechanisms of information transfer, this is the meaning attributed to it by Shannon in his development of information theory. It is \emph{not} used in the more general sense of implying a system of belief or theoretical framework as a context for that information. The same caveat should also be applied where we use the word `know' is the specific sense defined above.

From the definition of physicality given above, it immediately follows that if a `mind' can interact with a `body', then, \emph{by definition}, it is physical. Conversely, if a `mind' is \emph{not} physical, then it cannot interact with a `body'. Of course, this conclusion changes nothing at all regarding the actual nature of `minds' and `bodies': it has merely moved the linguistic goal posts. However, in order to speak meaningfully at all about philosophical issues, we must define our vocabulary rigorously. We shall argue that the above set of definitions is an extremely useful one to apply, since it clearly delineates the issues at hand.

\subsection{Mental concepts}
Let us consider the possible relationship between a non-physical mind and a physical body. How would the mind `know' anything about the world the body exists in? By definition, it does not interact with the body, which means, also by definition, that no information can be exchanged between mind and body. Hence, the mind cannot know that the body is walking down the road or taking a shower; nor can the body respond to the dictates of the mind to turn left at the corner or turn the hot tap on. Any such communication \emph{would} be an exchange of information, which we have defined to mean an interaction. The possibility of this means that both entities are \emph{physical} under the definitions given above.

If this idea leaves the reader feeling a little cold, it is very possibly due to the abiding prejudice within many people that associates `physicality' with the objects of our perceptions, rather than the interactions those perceptions are based on. That is, we tend to conceive of physical objects as existing independently of their interaction with other things, rather than being the manifestation of interactions that all things partake in.

Another powerful objection may lie in the subjective feeling of repugnance that our emotions and thoughts could be quantified in the same way as a pile of rocks. We are conscious, sentient beings with feelings and dreams. The idea of reducing that to some physical description is quite anathema to many. However, this emotive reaction is deeply rooted in the perception of what the `physical' is. It may be that the notion of what constitutes the physical world arising out of quantum theory may sit much more comfortably with the sense we have of our humanity.   

A commonly held belief is that the phenomenon of consciousness is `not physical' and, sometimes, that physical objects can not be conscious. The notion of consciousness is, of course, extremely difficult to define exactly but we may at least give one necessary criterion: that it must involve having information of some kind. We cannot be aware of anything if we have no information about it. This means that a conscious entity must interact with the objects of its awareness in some way. For now, we defer deeper discussion of this topic until Section~\ref{sec:quantum_minds}, after we have explored the relevant insights that quantum theory has to offer.

\subsection{Physical concepts}
Let us consider a fundamental concept that occurs both in our everyday perceptions and the majority of our most advanced physical theories: \emph{space}. This is a concept of fundamental importance, since many dualistic theories of mind and body use `spatial extension' as a criterion for defining what is physical. But what do we mean by space? Before considering what physics has to say, let us consider our everyday perceptions.

A common notion is that we perceive space all around us. In terms of a mental concept, this is quite true, but in terms of something that we can sense directly, this is clearly not the case. We cannot see, hear or touch space: our only sensory data is of the objects that we perceive to be `in space'. In other words, we infer the nature of space from the behaviour of objects that we sense directly. It is instructive to realise that this inference is therefore only possible on the basis of interactions between ourselves and these `objects'. 

After some reflection it should also become clear that neither do we directly sense `objects'. Rather, our perception of an object is inferred from many billions of interactions (photons interacting with electrons) that provide the input to our sensory nervous systems. These, in turn, communicate these data via electronic signals that are then extensively processed in our brains. Our notions of `objects' are then inferred via these conceptual processes. After some consideration, the reader may concur that these two types of inferences are very similar, since our notion of a physical object is very much related to the space it occupies.

Here is where the student of quantum mechanics becomes unstuck. When one first studies the theory, it seems that the world it describes is nothing like the world we experience at all. Rather than well defined `objects', we encounter `wavefunctions', smeared out over all of space. Entities do not sit obediently at one point in space as perceived objects do but seem to be `everywhere at once'. Moreover, the properties that `quantum entities' have do not seem to be fixed but depend on how we observe them.

Here we offer an insight: classical physics gives us a model of our \emph{perceptions} of reality, whereas quantum physics gives us a model of the reality encompassing the larger picture, describing how physical systems interact. The two are never going to seem to be commensurate, as they are describing different things. In order to understand how our perceived world emerges out of the larger quantum picture, we first need to compartmentalise the Universe into subsystems and then understand how information is transferred between them.

\section{Aspects of quantum theory}\label{sec:quantum}

\subsection{Superposition and uncertainty}
One of the most counter-intuitive concepts arising out of quantum mechanics is related to the \emph{principle of superposition}. A physical entity - we shall call it a `system' from here on - may be found in many different possible states. For instance, an elementary particle might be found at any number of different positions or with any range of momentum or energy. What quantum mechanics tells us is that the system has the potential to exist in \emph{all} these different possibilities simultaneously. Each possible state of the system (known as an `eigenstate') is weighted with a quantity called an `amplitude' (a complex number) and the total state is described by a superposition of all of these states. 

These amplitudes are associated with a very important interpretation known as the \emph{Born rule}~\cite{born1926quantenmechanik}. If we take the squared modulus of an amplitude, the resulting quantity is interpreted as giving the \emph{probability} for actually finding the system in the corresponding state. In a sense, it is the Born rule that transforms quantum mechanics from a body of pure mathematics into a physical theory. However, such an interpretation is not without problems, which we shall address in more detail in Section~\ref{sec:quantum_information}.

A further strangeness arises because any set (known as a \emph{basis set}) of eigenstates describing a particular physical property of the system is not unique. For instance, we might have one set of eigenstates describing all the possible values (known as \emph{eigenvalues}) of position but require a different basis set for describing the eigenvalues of momentum. Position and momentum are an example of what we call `conjugate variables'. An eigenstate of position is actually composed of a superposition of eigenstates of momentum and vice versa. 

This means we can never simultaneously measure both position and momentum with total accuracy. The more precisely we try to measure position, the more uncertain its momentum becomes and vice versa. This is known as the \emph{Heisenberg Uncertainty Principle}. The upshot of this is that the precise state of a system is rarely well-defined but can take a range of possible observable values.
 
This, of course, is totally at odds with the way we perceive the world. We experience things to be in a particular place at a particular time and not in some strange superposition of being everywhere at once. Moreover, we tend to think of the observable properties of a physical entity as being intrinsic to it; as something objective that does not depend on how we measure it. This, however, relates to our perception of the physical world, which is based on what happens when we interact with a system, not on its state of being when we do not.

\subsection{Coherence and decoherence}

Since quantum amplitudes are complex quantities, they have both magnitude and phase. Due to these phase factors, a superposition of quantum states  can exhibit \emph{interference phenomena}. This happens, for instance, when two amplitudes are pointing in the opposite direction (which we call being `out-of-phase') in which case they can cancel out, causing \emph{destructive} interference. On the other, when they are pointing in the same direction (`in-phase') they can add together causing \emph{constructive} interference. 

Interference phenomenon had long been known in optics ever since Thomas Young's famous demonstration of the wave-like properties of light in his `double slit' experiment~\cite{young1802ii}. In this, light is passed through two parallel slits in a screen before being projected onto a second screen beyond it. Because of the different path lengths from each slit to a particular point on the projection screen, the phases of the two light rays will be generally be different, causing either constructive or destructive interference. This results in a series of parallel light and dark bands, which disappear when one of the slits is blocked off. In 1927, the same diffraction phenomenon was demonstrated for electrons by G.P. Thomson~\cite{thomson1928experiments}, conclusively proving their wave-like nature. 

The fact that well-defined interference lines appear in the double-slit experiment is because the phase variation of the photon or other particle as it passes through the system remains regular and uninterrupted. Technically, we say that it remains \emph{coherent}, which means the phases at any point in time or space remain in a well-defined relation. If we were to disrupt this smooth variation of the phase in some way, the interference effects would disappear. This would occur, for instance, if we tried to measure which slit the particle went through by placing some kind of detector there. This interaction with the system typically means that information about it is transferred to its environment. This process is called \emph{decoherence}. Hereafter, wherever we speak of \emph{decoherence theory}, we are essentially referring to a theoretical model of how information is transferred from one system to another. 

\subsection{The measurement problem}
If a physical system really exists in a superposition of possibilities, why do we not perceive this? A traditional answer to this is that when we interact with a system, its quantum state `collapses' and changes discontinuously from a superposition of possibilities to the particular outcome that we perceive. It should be stressed, however, that this is \emph{not} part of the formal apparatus of quantum theory! This is an \emph{ad hoc} add-on to the theory and to date there is no universal consensus as to how, why or even \emph{if} this `collapse' occurs. The question of exactly what does happen and, in particular, what happens to all the other possibilities that we do not observe, is an aspect of what we call the \emph{measurement problem}. 

Interpretations of the theory that attempt to explain this may be divided into two categories that we can call \emph{subjective} and \emph{objective collapse}. Proponents of objective collapse argue that the quantum state does actually change discontinuously and that all the other possibilities are somehow destroyed. Many of these interpretations come under the umbrella of \emph{quantum mechanics with spontaneous localisation} (QMSL), the most well-known example being the Ghirardi-Rimini-Weber (GRW) model~\cite{ghirardi1986unified}. An important aspect of these interpretations is that they require a modification of quantum theory, introducing an as-yet-unknown collapse mechanism to interfere with the deterministic evolution of the quantum state.

In subjective collapse theories, on the other hand, it is held that there is \emph{no} collapse of the quantum state. Rather, the collapse is only perceived from the point of view of an observer who finds themselves in a `relative state' of the observed system. Such interpretations include the \emph{many worlds} interpretation~\cite{everett1956theory, dewitt2015many} and Zeh's \emph{many minds} interpretation~\cite{zeh1970interpretation, albert1988interpreting}. 

Given that, as yet, there is no established quantum theory of gravity, it is possible that some additional, non-linear mechanism causing objective collapse may be at work. A plausible candidate has been offered by Penrose~\cite{hameroff1996orchestrated, hameroff1998orchestrated} (which he calls \emph{objective reduction} or OR) based on the fact that in general relativity, a massive object will warp spacetime to some extent. A superposition of objects in different places would then correspond to a superposition of incommensurate spacetime geometries, which may result in some non-linear dynamical process selecting one option or the other.

For our purposes, we need not opt for one or the other of these broad interpretations. Both may be discussed in terms of decoherence theory and, as such, at the heart of each is the concept of \emph{information transfer} from one physical system to another, which we shall address in more detail in Section~\ref{sec:quantum_information}. The essential difference is that in subjective collapse theories, the total information content of the Universe is conserved, whereas in objective collapse it is not. This may well have consequences in the field of quantum information theory, where fundamental theorems such as \emph{no-cloning}~\cite{wootters1982single}, \emph{no-deleting}~\cite{pati2000impossibility} and \emph{no-hiding}~\cite{braunstein2007quantum} imply the conservation of information. We shall return to this issue in Section~\ref{sec:quantum_minds}.

\section{Quantum information}\label{sec:quantum_information}
\subsection{Probability and information}
Earlier, it was suggested that it is the Born rule that transforms quantum mechanics from an exercise in abstract mathematics into a physical theory. It is in the association of probabilities that we will find a system in a particular state that allows us to make practical predictions about the world. However, the notion of such probabilities is not unproblematic. To begin with, in statistics, there are at least two distinct ways to think about what a `probability' is. 

The first is the \emph{frequentist} interpretation. This says that, given a number of repeated, identical trials, the frequency of a particular result divided by the total number of trials approaches the \emph{probability} for that particular outcome. Whilst this does imply that there may be some objective criteria determining how often a particular event occurs, it remains an empirical prescription that cannot be unequivocally associated with a particular state, which may be measured only once.

In contrast to this, there is the \emph{Bayesian} approach, that says that a probability is no more than a statement of belief about the nature of a system. The belief is that a particular event has, \emph{a priori}, a certain chance of occurring and it is the job of Bayesian analysis to set up an iterative process for refining our beliefs. The problem with this approach is that a `belief' is generally something subjective and so does not necessarily apply objectively to the system in question. Although, having said this, belief does imply something objective that is `believed in'.

A third notion arises out of classical information theory. In his analysis of a transmitted message consisting of binary digits (`bits'), Claude Shannon~\cite{shannon1948mathematical,shannon2001mathematical} came up with a mathematical definition of the \emph{information content} of a sequence of tokens.  He defined a unit of information in terms of the probability for the occurrence of a particular token in the sequence. He then called the weighted average of these terms the \emph{entropy}, due to its mathematical equivalence with the Boltzmann-Gibbs expression for thermodynamic entropy~\cite{boltzmann1866mechanische, jaynes1965gibbs}. At the time, many thought this resemblance to be purely accidental, although with the hindsight of modern quantum information theory, we may now interpret thermodynamic entropy in informational terms (more on this in the Section~\ref{subsec:entropy_information}).

For our purposes, we have already \emph{defined} the quantum state of a system to represent its informational content. So we suggest an alternative view: that we consider \emph{information} to be primary and that the \emph{probability} for a particular state may be extracted from it. The concept of probability that emerges remains consistent with both the frequentist and Bayesian interpretations but has its ontological roots in the notion of information.  

\subsection{Entropy and information}\label{subsec:entropy_information}

The thermodynamic concept of \emph{entropy} is often described as a measure of `the disorder' of a system. Whilst entropy \emph{is} related to disorder, this definition rather lacks in utility and, more importantly, fails to encapsulate the truly profound nature of the concept. To cut a long story short, entropy is a \emph{measure of information}. Very often, it may be used in the negative sense as a \emph{loss} of information but mathematically, this difference lies in the presence or not of a minus sign.

Perhaps the first person to highlight this connection was James Clerk Maxwell with the invocation of his now notorious `demon'. This was a creature possessed of total knowledge of a system who could violate the Second Law of Thermodynamics by reducing the entropy of a system and hence extract work from it. 

Boltzmann further developed the concept of entropy by introducing a probabilistic form for it~\cite{boltzmann1866mechanische}. Essentially, he described the entropy of a macroscopic system as the logarithm of the number of possible microscopic configurations that it could occur in. 

The connection to `disorder' is then fairly easy to see. Consider the example of a box containing a number of particles, where, initially all the particles are constrained to be in one half of the box by a partition. We may view this as a more `ordered' state of affairs than one in which the particles could be found anywhere in the box. Once the partition is removed, the particles are free to diffuse throughout the container. There are now a far greater number of possible configurations for the particles to be in. Hence the entropy has increased and the system is in a more `disordered' state.

What is truly significant about this though, is the notion that the entropy of the system \emph{increases} with time, giving temporal evolution a definitive direction (often referred to as the `arrow of time'). This is essentially what the Second Law of Thermodynamics states. This, however, is completely at odds with all other laws of physics,\footnote{By `laws of physics', we mean here the mathematical models that we have constructed to explain how Nature works.} which are all \emph{time symmetric}. They have no preferred temporal direction and work just as well if we take time to be running backwards as forwards. In other words, the laws of physics are \emph{reversible}. How, then, do we reconcile this with the Second Law, which clearly asserts irreversibility?

One answer is to argue that the more entropic states have a greater probability, so the system is more likely to evolve into them. However, probability ought to play no part in deterministic laws. Given the state of a system at a given time, under deterministic laws it will evolve into another, well-defined system at a later time \emph{with certainty} - there is no probability involved! At the time Boltzmann published his work, this objection was framed as the \emph{Loschmidt paradox}~\cite{wu1975boltzmann}, which asserted that irreversible dynamics \emph{cannot} emerge out of reversible physical laws. 

The essence of a solution is that the apparent increase in entropy of the Universe occurs when there is a transfer of information from one system to another. We shall see that this depends on a quantum mechanical phenomenon known as \emph{entanglement}, which we shall describe in Section~\ref{subsec:entanglement}. It turns out that entanglement \emph{requires} interaction between the two systems, which highlights the fact that entanglement, interaction and information transfer are all fundamentally related.

At the Universal level, this process of entanglement is still reversible. That is, entangled systems may become disentangled. However, it may be argued that from the perspective of an entity able to store information robustly, information accrual must necessarily be perceived as an irreversible process, as we shall discuss further in Section~\ref{subsec:info_memory_irreversibility}. Hence, the time-asymmetry inherent in the Second Law of Thermodynamics may be seen in terms of the accrual of information at intervals of `entropic time' steps~\cite{vaughan2020concept}.

\subsection{Entanglement}\label{subsec:entanglement}

\begin{figure}[!t]
\centering
\includegraphics[width=0.75\textwidth]{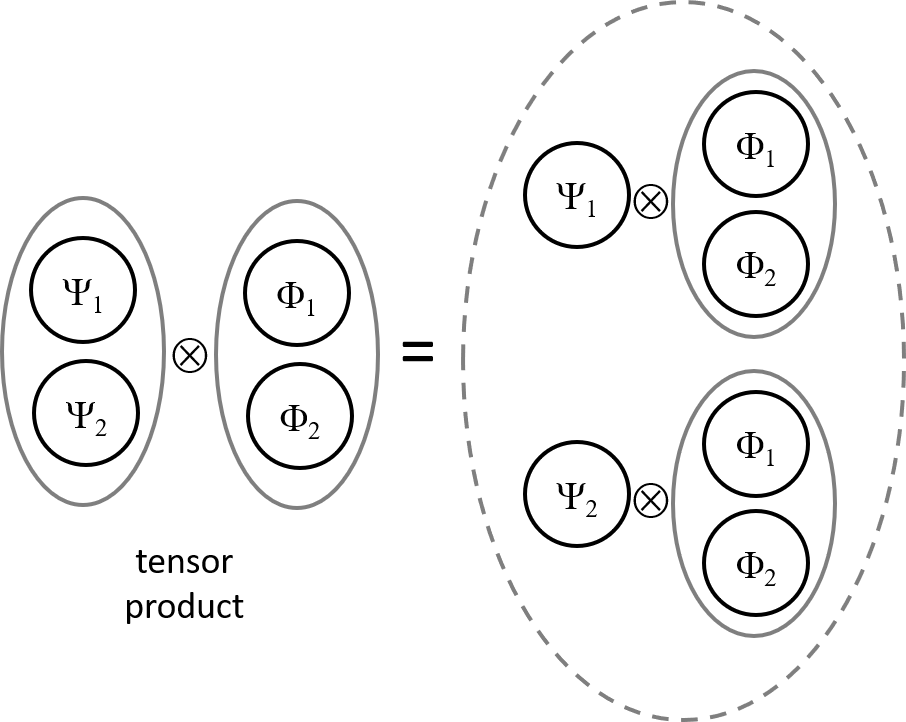}
\caption{Schematic of the \emph{tensor product} of two systems. This may always be factored into a product of the isolated systems as shown on the left.}\label{fig:tensor_product}
\end{figure}

Entanglement is a uniquely quantum mechanical phenomenon, providing the mechanism by which a correlated exchange of information between two physical systems can occur. Let us suppose that we have two systems, which we shall label $\Psi$ and $\Phi$ for ease of reference. For simplicity, let us suppose that each system is only two-dimensional, meaning that for a particular choice of basis set, each only has two possible eigenstates, which we shall label with subscripts $1$ and $2$. 

In the absence of any interaction between the two systems, the total composite system may be represented by a \emph{tensor product}, illustrated schematically in Fig.~\ref{fig:tensor_product}. In this case, each component of either system is multiplied by each component of the other, meaning that for the two-dimensional systems considered here, the total system would have four components altogether. However, rather than show these four components explicitly, in Fig.~\ref{fig:tensor_product}, we illustrate how this is equivalent to each component of system $\Psi$ multiplying the \emph{entire} system of $\Phi$ (or vice versa). This makes it clear that the components of either system are independent of those of the other.

Consider for instance, the case in which $\Psi$ is found to be in the state $\Psi_{1}$. From the figure, we see that $\Phi$ may be in either state $\Phi_{1}$ or $\Phi_{2}$. In other words, \emph{knowing the state of $\Psi$ gives us no information about the state of $\Phi$} (and vice versa).

\begin{figure}[!t]
\centering
\includegraphics[width=0.75\textwidth]{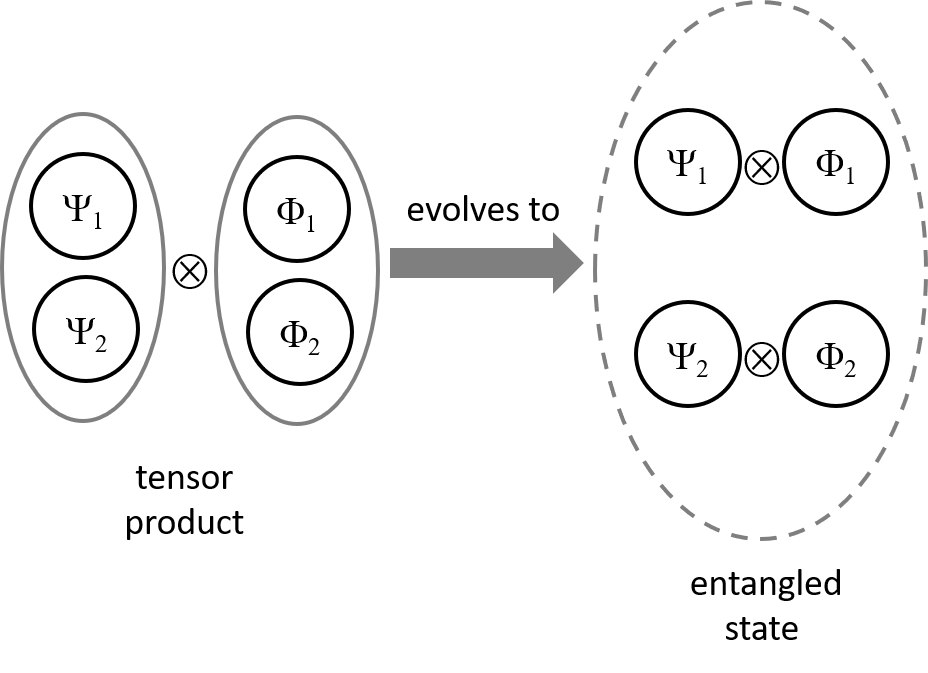}
\caption{Schematic of \emph{entanglement} between two systems. The essence of an entangled system is that it \emph{cannot} be factored into a product of the isolated systems.}\label{fig:entangled_state}
\end{figure}

Compare this with the situation depicted in Fig~\ref{fig:entangled_state}, in which some interaction between $\Psi$ and $\Phi$ causes the total system to evolve into the state on the right. Here, $\Psi_{1}$ is coupled \emph{only} with $\Phi_{1}$, whilst $\Psi_{2}$ is coupled \emph{only} with $\Phi_{2}$. This means that if we were to find $\Psi$ in state $\Psi_{1}$, we would certainly find $\Phi$ is state  $\Phi_{1}$. In other words, \emph{information about the state of $\Phi$ has been encoded into the $\Psi$} (and vice versa). This is an example of \emph{entanglement}. Note that the defining feature of entanglement is that the entangled state \emph{cannot} be factorized into a tensor product, such as in Fig.~\ref{fig:tensor_product}.  

The example shown in Fig~\ref{fig:entangled_state} illustrates the case of a \emph{maximally} entangled system. More generally, the degree of entanglement may be less than this. For instance, instead of knowing that $\Phi$ is in $\Phi_{1}$ if $\Psi$ is in $\Psi_{1}$ with \emph{certainty}, we may just have the case where there is an enhanced probability for this to be the case (over and above the intrinsic probability for this when we just consider $\Phi$ in isolation). There are many ways to quantify this but for our purposes the most relevant is a quantity known as the \emph{entropy of entanglement}, which may be calculated using the methods of decoherence theory. This may be interpreted as a measure of the total information transferred between the two systems.

\subsection{Relative states}

\begin{figure}[!t]
\centering
\includegraphics[width=0.75\textwidth]{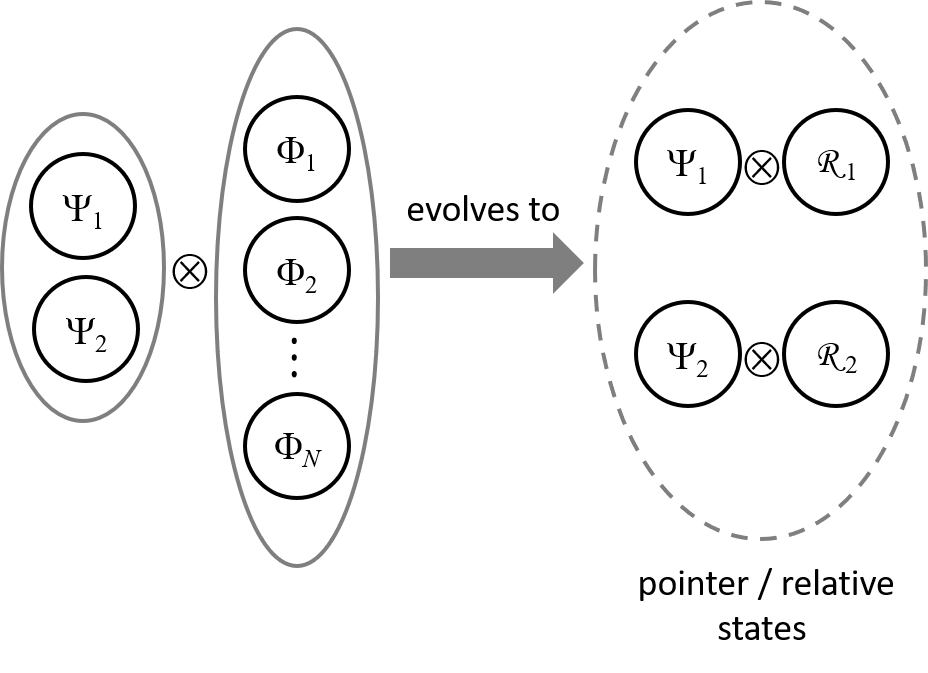}
\caption{Schematic of premeasurement in terms of \emph{pointer} and \emph{relative} states.}\label{fig:pointer_relative}
\end{figure}

Let us extend the example of Section~\ref{subsec:entanglement} to the particular case where $\Psi$ is some `observed' system and $\Phi$ is its environment with which it interacts. To make this realistic, we must greatly increase the dimensionality of $\Phi$, as illustrated in Fig.~\ref{fig:pointer_relative}. As $\Psi$ interacts with its environment, we see a similar picture to Fig.~\ref{fig:entangled_state} emerging except that now the components of $\Psi$ are now associated with states $\R_{1}$ and $\R_{2}$, which are constructed from \emph{superpositions} of the $\Phi_{i}$ states. These are known as the \emph{relative states} of the $\Psi$ states. 

Figure~\ref{fig:pointer_relative} illustrates the scenario of \emph{premeasurement}, in which the interaction between $\Psi$ and its environment as picked out certain \emph{pointer states}~\cite{zurek1981pointer} and associated these with \emph{relative} states~\cite{everett1956theory}, which encode information about the pointer states. It is called `premeasurement' because the process has \emph{not} selected out \emph{which} of these pointer states we actually observe. In fact, whilst decoherence theory can explain which particular \emph{set} of states are picked out (the `preferred basis'), it still does not answer the `which state' question.

\subsection{Information, memory and irreversibility}\label{subsec:info_memory_irreversibility}
It should be noted that there is no \emph{a priori} reason why information transferred between systems should remain intact as the total system evolves. Quantum mechanics is what is known as a \emph{unitary} theory, which essentially means that anything that is done can be undone, or in technical jargon is \emph{reversible}. This means that systems that become entangled may become disentangled at some other time, losing any information that has been transferred.

However, it is also quite possible (and very often the case) that information may be stored robustly in a relative state. Such storage will then constitute \emph{memory}. As the system evolves, there may be many possible states in which this information is added to and perhaps countless others in which it is erased. However, from the point of view of any observer embedded in a relative state (i.e. they are a part of it), only states containing a `history' will give any sense of evolving time. 

Imagine for, example, information $I_{1}$ about an event at time $t_{1}$ is stored in a relative state in some way, which is then supplemented by some additional information $I_{2}$ at a different time $t_{2}$. If the system then evolves (deterministically) in some way such that $I_{2}$ is then erased, the observer is then effectively thrown back in time. The observer, quite literally, has no information about the event at $t_{2}$. From the observer's point of view, this event would not have `happened'.

The only thing an observing system can be aware of is the information embedded in the relative state it is also a part of. From its subjective point of view, then, this accrual of information is necessarily \emph{irreversible}. Moreover, the order of the time steps $t_{1}, t_{2}, \ldots$ associated with this accrual of information need not follow the temporal parameter `$t$' representing time in the deterministic formulation of the theory. The only requirement is that units of information be consistently related. Specifically this means that subsets of information representing `early memories' are being contained within larger sets constituting `later memories'. An initial formulation for such a scheme of things may be found in Ref.~\cite{vaughan2020concept}, where these ideas about `entropic time' are explored in more mathematical detail.

Before leaving this section, it is relevant to note that under an objective collapse interpretation, such irreversible accrual of information is guaranteed in all cases (not just the ones it is possible to `remember'). Once the quantum state has collapsed it cannot `uncollapse', since all information about other possible states has been lost. These states cannot then conspire to erase information stored in the remaining relative state. 

\section{Quantum minds}\label{sec:quantum_minds}
\subsection{Consciousness}
In the description of information and irreversibility in Section~\ref{subsec:info_memory_irreversibility}, we deliberately abstained from use of the word `conscious', when describing an observer. This is appropriate for a general discussion of quantum information since in the broadest sense, an `observer' could be any system able to receive information. Our discussion did, however, specialise somewhat to the case of an observer able to store information. This is of particular relevance in the case of conscious entities such as ourselves, since our existential reality is composed in large part of our memories and psychological experience of time.

In this section, we specifically focus on the case of conscious observers to enquire how the ideas from quantum information theory we have been exploring might apply to the concept of mind. Here, it is important to emphasise that we do not seek to `explain consciousness' in terms of some more fundamental concept. Indeed, whilst one may point to correlations between neural processes and conscious experiences, the question of \emph{why} such processes should manifest in conscious experience has always escaped explanation. This is has been referred to by Chalmers as the `hard problem' of consciousness \cite{chalmers2010character} - how can we explain the experience itself?


The intractability of the `hard problem' has lead some investigators to the view that we should take conscious experience to be fundamental and therefore not explicable in terms of anything more primary. This is the view taken by Tononi~\cite{tononi2015integrated, oizumi2014phenomenology} in the development of an approach known as \emph{integrated information theory} (IIT), with which the present work shares many commonalities. In particular, IIT focuses on the informational content of conscious experience.

\subsection{Informational content of consciousness}
On consideration of our own experiences, it is evident that states of consciousness have considerable information content. This follows from the fact that, as Tononi points out, conscious experience may be differentiated into specific experiences. As a highly simplified example, consider the state of consciousness of an individual only able to differentiate between light and dark. These two elementary experiences may then represent one `bit' (binary digit) of information.\footnote{More generally, one might choose other units of information for an individual capable of experiencing all intermediate shades of gray.}

Our actual experiences are vastly more complex than this, involving many different types of experience. Here, it is the different subjective experiences of qualia (e.g. tones, textures, colours, tastes) that allows such a high degree of information to be encoded. These are the very experiences that defy functional explanation (constituting the `hard problem') and yet, in themselves, they provide a functional capacity - they allow consciousness to constitute an informational state. By the definition of physicality that we are endorsing in this paper, conscious experience is therefore a \emph{physical} state.

If we are to accept that conscious states are physical (which we are saying is the case by definition), this does then beg the question as to whether the converse is true: are all physical states conscious? Such a question must be tempered by the appreciation that the `consciousness' of a physical system only capable of supporting a single bit of information would be a truly unremarkable experience. It would be unable to support memory or represent any detailed picture of the world it existed in. Such a reality would be wholly unlike the conscious experience enjoyed by human beings. The degree of information that could be represented, and hence the information that could be exchanged with other systems, is clearly a significant factor.

Quantifying the information content associated with consciousness is one of the primary objectives of the IIT project. However, whilst IIT makes great use of Shannon's information theory, there appears to be little mention of its generalisation to quantum information theory\footnote{To the author's knowledge.}. The question naturally arises: how does a system of composite parts obtain some irreducible level of information content? That is, why is any system not reduced to a collection of simple systems only capable of supporting a very small amount of information? The natural answer to this arising out of quantum theory is \emph{entanglement}, which naturally gives rise to such irreducibility.

As attractive as the invocation of entanglement to solve this problem is, it is not without problems itself. The entangled states deliberately sought in quantum computing are often fragile, being highly susceptible to decoherence with the environment. Moreover, the question of whether coherent states can be maintained in neural systems has been called into severe doubt~\cite{tegmark2000importance} (as discussed later). On the other hand, entanglement with the environment is the very process by which information is communicated between systems. The question as to whether consciousness may be associated with a robust, entangled state therefore remains open. 

\subsection{Characteristics of consciousness}
Having argued for the physical nature of consciousness, in this section we take an overview of the phenomena of consciousness, arguing for the consistency of this stance. Let us first assert some key characteristics of conscious experience, before considering each in more detail:

\begin{itemize}
\item Awareness (knowledge of the world)
\item Holism (connectedness of experience)
\item Temporal localisation (sense of `now')
\item Memory (sense of `past')
\item Personal identity (lack of omniscience)
\end{itemize}

\noindent These categories are proposed as being relevant in the present context of this paper and the concepts discussed therein. They are intended to be descriptive of conscious phenomena, rather than indicative of axiomatic truths. Each of these does have some commonalities with the axioms of IIT but, with the exception of `Holism', there is no one-to-one correspondence.

\subsubsection{Awareness}
Awareness is, in one sense, the easiest of all these characteristics to give a quantum mechanical description to. The meaning of the word implies `knowledge of', which we have described extensively in terms of information transfer. This, in turn, has been explained in terms of quantum entanglement.

What remains unexplained is the qualitative \emph{experience} of awareness. This is the nature of consciousness itself and its explanation constitutes Chalmer's `hard problem'. If fact, we would argue that qualitative experience \emph{cannot} be explained, since a functional explanation requires conceptual handles to encapsulate an idea. Only the \emph{substance} of that experience (which amounts to information) can provide such grabbing points. 

We can, however, argue that nothing about this experience is inconsistent with a quantum mechanical explanation, so nothing metaphysical need be invented to explain it. Indeed, any metaphysical explanation would be just as impotent at explaining qualitative experience, for the same reason given above.

\subsubsection{Holism}
One of the characteristic features of consciousness is that of experiencing many different things at once. At any point, we may be aware of sights, sounds, smells and touch as well as our inner thoughts and feelings. These all seem to come together as a single, holistic experience. This phenomenon has, in fact, been taken as one of the axioms of IIT under the name `integration'~\cite{tononi2015integrated, oizumi2014phenomenology}.

We might then ask how it is that we can be aware of so many different aspects of the world simultaneously, given that, classical physics at least, is premised on the notion of `local causes'. That is, a state is only affected by what it is immediate contact with.

A consistent framework for this is quantum entanglement as suggested earlier. We have already seen how information may become transferred from one system to another via entanglement. Once entangled, the system can no longer be viewed as the sum of its parts - it is essentially a single state. In line with our description of quantum mechanics as a theory of information, the state itself represents the information encoded into it, which is multitudinous.

An open question at this point is the degree of coherence of a conscious state. Hameroff and Penrose~\cite{hameroff1996orchestrated, hameroff1998orchestrated} argue that the brain may sustain in coherent `pre-conscious' state before collapsed due to their proposed `OR' mechanism, giving rise to consciousness. This idea has been criticised~\cite{tegmark2000importance} on the basis of the extremely short decoherence time that neural phenomena would experience, suggesting that no such coherent state could persist. 

\subsubsection{Temporal localisation} 
A significant feature of our conscious experience is the sense of existing in a particular `now' moment. On the face of it, this would seem to have no explanation in the context of time-symmetric laws of physics in which no particular point in time is singled out as being special. 

From the point of view in which temporal evolution is seen as the accrual of information, however, this is unproblematic. All a conscious observer can ever be aware of is the sum total of information he or she has access to. Future events are events that the observer has no information about and so cannot be encompassed in a state of awareness.

Whilst this description certainly explains our sense of the past, it is noteworthy that our memories are never as rich and vivid as the immediate experience of our world. Walking through a wood, for instance, we are subjected to a deluge of simultaneous experiences: the texture of the bark of trees; the sounds of birds and insects; the feel of the air on our faces and so on. These experiences represent a huge amount of information that our brains are able to store only ephemerally. After the fact, only aspects of these experiences may be stored as longer term memories (the recollection of which also contributes to our immediate sense of awareness). 

The present moment is indeed special - it represents the point at which we are storing a huge amount of fleeting information. This \emph{is} an informational state but not one that can endure. Hence, by linking our information about our current experiences with recent ones, we have a sense of moving from one moment to the next.  

\subsubsection{Memory}
The creation of our sense of the `past' via stored information (memories) has already been extensively been discussed throughout this paper. Here, we merely reiterate that such memory is the natural consequence of storing information that arises, in the first place, through physical interaction. 

\subsubsection{Personal identity}
Our sense of personal identity must be predicated on some sense of ourselves, which in turn requires that we have stored information in the form of memories. However, the notion of personal identity goes further than this and requires that we are distinct in some way from the rest of the Universe.

Suppose we were aware of \emph{all} things; of \emph{everyone's} thoughts and feelings. How could we then define personal identity? Nothing would distinguish one set of experiences from another. Our sense of personal identity can therefore only come about due to the fact that we are \emph{not} omniscient and that our experience of the world is filtered down to a restricted set of data. 

This is readily explained in terms of the limited capacity of a system to store information (it is fundamentally limited by its degrees of freedom). What identifies a particular system as being `individual' is a somewhat harder problem. Where exactly is the division between one thing and the next? This is a tough question to pin down technically but we may argue from a high-level point of view that an individual must be defined in informational terms.\footnote{As mentioned earlier, such quantification is one of the objectives of the IIT program, although a review of this is beyond the scope of this paper.}

Specifically: what information does an individual have access to that another does not? This is most obviously answered in terms of thoughts, feelings and sensory awareness. Such processes correspond to physical systems and the holistic experience of these may correspond to a state of entanglement between them. 

\subsection{The conservation of information}
Although we have endeavoured to remain neutral on the `which state' aspect of the measurement problem, we did indicate earlier that this has implications for the conservation of information. This is the idea that the informational content of the entire Universe is never depleted or added to, even though although it may appear to be lost at a local level, manifesting as an increase in entropy. The conservation of information is implied by three theorems of quantum information theory:

\begin{itemize}
\item An arbitrary quantum state cannot be cloned (we cannot create information)
\item An arbitrary quantum state cannot be deleted (we cannot destroy information)
\item Information lost from one system is transferred to another (information cannot be hidden from the Universe)
\end{itemize}

\noindent These are known as the \emph{no-cloning}~\cite{wootters1982single}, \emph{no-deleting}~\cite{pati2000impossibility} and \emph{no-hiding}~\cite{braunstein2007quantum} theorems respectively. Note that these are \emph{theorems}, not theories. That is, if the postulates of quantum mechanics are correct, then they are \emph{necessarily} true.  

Of course, it is quite possible to clone and delete \emph{classical} information without any problem. However, in terms of bits, these are either 1 or 0, whereas a quantum bit (a \emph{qubit}), may be in a superposition of 1 and 0 and it is the amplitudes of these states that carry the information. Moreover, even in classical theory, the erasure of a bit of information is known to transfer a certain amount of energy to other degrees of freedom of the environment, representing an increase in entropy. This is known as \emph{Landauer's Erasure Principle}~\cite{landauer1961irreversibility, landauer5dissipation}. Quantum mechanically, via the no-hiding theorem, this lost information is just transferred to another part of the environment - it is never lost.

The problem with objective collapse theories is that they violate the conservation of information. If a quantum state collapses discontinuously to a particular outcome, with the concomitant erasure of all other possibilities, then \emph{information is irreversibly lost}. Not only does this mean that nature is non-unitary, it follows that \emph{the postulates of quantum mechanics must be incorrect} (as they stand). This, of course, may well be the case but so far the theory has served us extraordinarily well.

\section{Conclusions}
The aim of this paper has been to elucidate aspects of modern quantum theory and thereby challenge popular notions of the meaning of the word `physical'. Throughout, we have adopted the view point that the true subject matter of a physical theory is information and that the dynamics of information transfer are as relevant to the discussion of `minds' as they are of `bodies'. To a large extent, this is a problem of language. It has therefore been necessary to clearly define what we mean by the word `physical' as well as what we mean when we use the word `know'. 

Given the definitions (1)-(5) of Section~\ref{sec:definitions}, it follows that a `mind' can only know of a `body' and vice versa if both are physical. If such things as `non-physical' minds existed, then they could know nothing of our physical state or have any influence on it. Hence, if our consciousness involves an awareness of the physical world around us, then that consciousness must be associated with a physical entity and can have nothing to do with a hypothetical non-physical mind.

We have then shown how these definitions find realisation in the theoretical models of quantum theory. Of especial importance is the phenomenon of entanglement, which provides the mechanism of information transfer. Here, the concepts and tools of modern decoherence theory are invaluable for the insights they provide.

It has been argued that nothing in our qualitative description of consciousness requires an alternative metaphysical explanation. Whilst not attempting to `explain' consciousness, its defining characteristics remain describable by physical concepts.

Finally, we must note a particular omission. We have said nothing on the matter of `free will', which for many is also a core characteristic of human ontology. The reason for this omission is simply the magnitude of the subject. Like the word `physical', we must first define what we mean by `free will', which is a major task in itself. Indeed, we would argue that the contention between compatibilists (those that claim free will is compatible with determinism) and incompatibilists (those that claim it is not) comes down to different definitions of what is meant by the term. The question is therefore left as the subject of further investigation.



\end{document}